\documentclass[twocolumn,prd,showpacs,amsmath,amssymb,superscriptaddress]{revtex4}

\usepackage{graphicx}
\usepackage{url}
\usepackage{dcolumn}
\usepackage{bm}
\usepackage{pgf,pgfarrows,pgfnodes,pgfautomata,pgfheaps,pgfshade}

\begin{document}
\title{Upper limit on the flux of photons with energies above
  $10^{19}$~eV using the Telescope~Array~surface~detector}

\author{T.~Abu-Zayyad}
\address{High Energy Astrophysics Institute and Department of Physics
  and Astronomy, University of Utah, Salt Lake City, Utah, USA}
\author{R.~Aida}
\address{University of Yamanashi, Interdisciplinary Graduate School of
  Medicine and Engineering, Kofu, Yamanashi, Japan}
\author{M.~Allen}
\address{High Energy Astrophysics Institute and Department of Physics
  and Astronomy, University of Utah, Salt Lake City, Utah, USA}
\author{R.~Anderson}
\address{High Energy Astrophysics Institute and Department of Physics
  and Astronomy, University of Utah, Salt Lake City, Utah, USA}
\author{R.~Azuma}
\address{Graduate School of Science and Engineering, Tokyo Institute
  of Technology, Meguro, Tokyo, Japan}
\author{E.~Barcikowski}
\author{J.W.~Belz}
\author{D.R.~Bergman}
\author{S.A.~Blake}
\author{R.~Cady}
\address{High Energy Astrophysics Institute and Department of Physics
  and Astronomy, University of Utah, Salt Lake City, Utah, USA}
% ...
\collaboration{Telescope Array Collaboration}

\begin{abstract}
We search for ultra-high energy photons by analyzing geometrical
properties of shower fronts of events registered by the Telescope
Array surface detector. By making use of an event-by-event statistical
method, we derive upper limits on the absolute flux of primary photons
with energies above $10^{19}$, $10^{19.5}$ and $10^{20}$\,eV
based on the first three years of data taken.
\end{abstract}

 
\maketitle

\section{Introduction}
The Telescope Array~(TA) experiment~\cite{Tokuno:2011} is a hybrid
ultra-high energy (UHE) cosmic ray detector covering about 700~km$^2$ in 
central Utah, USA. 
It is composed of a Surface Detector (SD) array
and three Fluorescence Detector (FD) stations.  
The TA SD array consists of 507 plastic scintillator detectors on a 
square grid with 1.2~km spacing~\cite{TASD}.  
They each contain two layers of $1.2$ cm thick plastic scintillator
3~m$^2$ in area.  
The three FD stations~\cite{TAFD} contain a total of 38 telescopes 
overlooking the air space above the array of scintillator detectors. 
The purpose of this paper is to present the
photon search capabilities of the Telescope Array surface detector and to
search for primary photons in the cosmic ray flux. We place the limits
on the integral flux of photons for energies greater
than $E_0$, where $E_0$ takes values $10^{19}$, $10^{19.5}$ and $10^{20}$\,eV.

At present there is no experimental evidence for primary UHE photons.
However, several limits on the photon flux have been set by independent experiments.  
These include Haverah Park~\cite{HP_lim},
AGASA~\cite{AGASA_1stlim}, Yakutsk~\cite{Ylim,Ylim18} (see also
reanalyses of the AGASA~\cite{AGASA_Risse} and
AGASA$+$Yakutsk~\cite{A+Y} data at energies greater than $10^{20}$~eV) and the
Pierre Auger Observatory~\cite{Auger_fdlim,Auger_sdlim,Auger_hyblim2}.

Photon limits may be used to constrain the parameters of top-down
models~\cite{Berezinsky:1998ft}.  The photon searches may be
used to assess parameters of astrophysical sources in the
Greisen-Zatsepin-Kuzmin~\cite{g,zk} cut-off scenario which predicts
photons as ever present secondaries. If UHE photons are observed, they
will be a supporting evidence for the GZK nature of the spectrum break at
the highest energies observed by HiRes~\cite{Abbasi:2007sv}, Pierre
Auger Observatory~\cite{Abraham:2008ru} and TA~\cite{TAspec}.  Photon
flux is sensitive to the mass composition of cosmic rays and hence
may be used as a probe of the
latter~\cite{Gelmini:2007jy,Hooper:2010ze}. The results of the photon
search also constrain parameters of Lorentz invariance
violation~\cite{Coleman:1998ti,Galaverni:2007tq,Maccione:2010sv,Rubtsov:2012kb,Satunin:2013an}.
Finally, photons with energies greater than $\sim 10^{18}$~eV could be
responsible for CR events correlated with BL Lac type objects on an
angular scale significantly smaller than the expected deflection for
protons in cosmic magnetic fields. This suggests neutral
primaries~\cite{Gorbunov:2004bs,Abbasi:2005qy} (see Ref.~\cite{axion}
for a possible mechanism).

Since the TA detectors are composed of thin scintillators, 
they respond equally to the muon and
electromagnetic components of the extensive air shower and are 
therefore sensitive to showers induced by photon primaries~(see e.g.\
Ref.~\cite{Nih} for discussion). We use the shower front curvature as
a Composition-sensitive parameter (C-observable) and a
modification of an event-by-event statistical method~\cite{ec_method} to
constrain the photon integral flux above the given energy. For the
Energy-sensitive parameter (E-observable), we use the scintillator
signal density at 800\,m core distance~$\mathcal S \equiv S_{800}$. The
comparison of an event-by-event statistical method with the ``photon median''
method~\cite{Auger_sdlim} is presented.

\section{Simulations}

\begin{figure}
\includegraphics[width=\columnwidth]{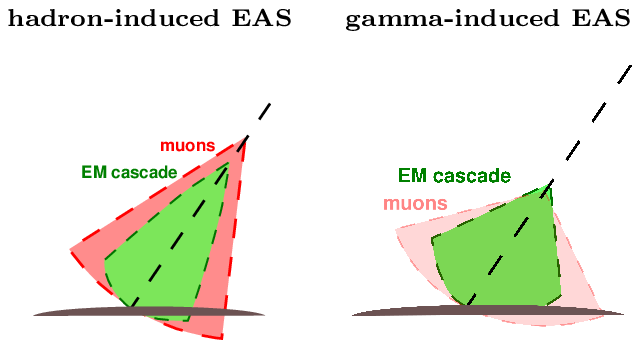}
\caption{\label{fig:front} Illustrative view of hadron- (left) and
gamma-induced (right) showers. Gamma-induced shower is deeper due to smaller
cross-section of the first interaction. Moreover, the hadronic cascade
is secondary with respect to electromagnetic in photon-induced
showers. The latter contains fewer muons (shown in red) and have
 larger curvature of the shower front. }
\end{figure}

Extensive Air Showers~(EAS) induced by photon primaries differ
significantly from hadron-induced events 
(see e.g.~\cite{RisseRev} for a review). 
Photon induced showers contain fewer muons and have 
a deeper shower maximum when compared to hadronic showers.  
The latter results in the shower front having more
curvature at the surface as illustrated in Figure~\ref{fig:front}.  
At the highest energies, 
there are two competing effects responsible for the
diversity of showers induced by photon primaries.  
First, the electromagnetic cross-section is suppressed at energies,
$E~>~10^{19}$~eV due to the 
Landau, Pomeranchuk \cite{LP} and Migdal \cite{M}~(LPM) effect. 
The LPM effect delays the first
interaction so that the shower arrives at ground level underdeveloped.
The second effect is $e^\pm$ pair production which is due to photon
interaction with the geomagnetic field above the atmosphere.  
Secondary electrons produce gamma rays by synchrotron radiation
generating a cascade in the geomagnetic field. The probability of
photon conversion is a function of photon energy and the perpendicular
component of geomagnetic field~\cite{Erber}. 
The shower development therefore depends on both the zenith and 
azimuthal angles of the photon arrival direction.

\begin{figure*}[t]
\includegraphics[width=1.0\textwidth]{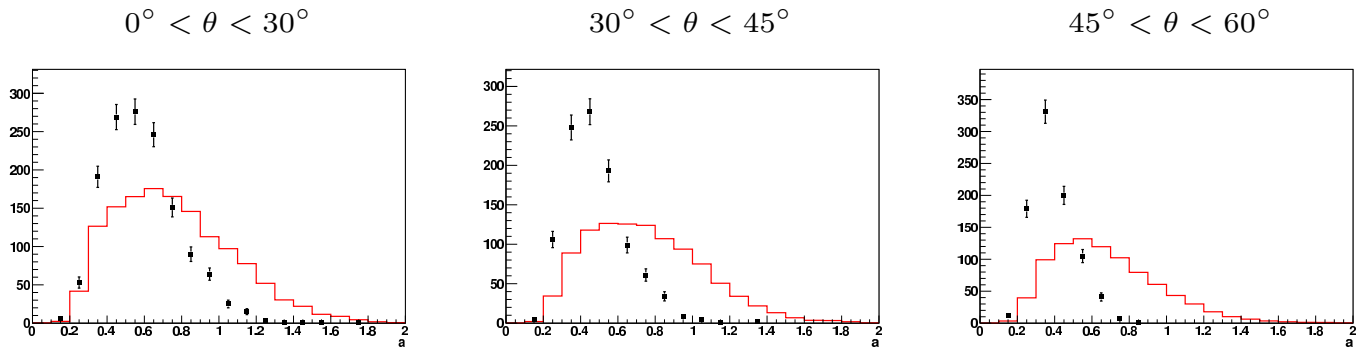}
\caption{\label{fig:a} Linsley curvature parameter distribution for
  three different zenith angle regions for reconstructed $E_\gamma >
  10^{19}$\,eV. The black points refer to data, red line represents
  the photon MC generated with an
  $E^{-2}$ spectrum.}
\end{figure*}

The event-by-event method~\cite{ec_method} requires a set of simulated
photon-induced showers for the analysis of each real shower.  We
simulate the library of these showers with different primary energies
and arrival directions.  For the highest energy candidates (events
which may be induced by a photon with primary energy greater than
$10^{19.5}$\,eV) we simulate individual sets of showers with fixed
zenith and azimuthal angles. At these energies, the shower development
becomes azimuth-angle dependent due to the photon cascading in the
geomagnetic field~\cite{RisseRev}.

We use CORSIKA~\cite{corsika} with EGS4~\cite{Nelson:1985ec} to model
the electromagnetic interactions and PRESHOWER
code~\cite{Homola:2003ru} for geomagnetic interactions.  There is no
significant dependence on the hadronic model because only
photon-induced simulated showers are used in the method.  The showers
are simulated with thinning and the dethinning procedure is
adopted~\cite{dethin} to simulate realistic shower fluctuations.

The detector response is accounted for by using look-up tables generated 
by GEANT4~\cite{GEANT} simulations. 
Real-time array status and detector calibration information are used 
for each Monte Carlo~(MC) simulated event. 
The Monte-Carlo events are recorded in the same format as real
events and analysis procedures are applied in the same way to both. 
The photon-induced MC set contains $2\times 10^{6}$ triggered events
produced from 3380 CORSIKA showers by randomizing core location~\cite{StokesMC}.

\section{Data Set}

We use the Telescope Array surface detector data set observed and
recorded between 2008-05-11 and 2011-05-01. 
During this time period, the surface detector array
was collecting data with a duty cycle greater than 95\%~\cite{TASD}. 

We reconstruct each event with a joint fit of the geometry and Lateral
Distribution Function (LDF) and determine the Linsley curvature parameter
``$a$'' (see Appendix~A for definition) along with the arrival
direction, core location, and signal density at 800 meters $\mathcal S
\equiv S_{800}$. 
As noted above, the same reconstruction procedure is applied to both
data and Monte-Carlo events.

For each real event, ``$i$'', we estimate the energy of the hypothetical
photon primary, $E^i_\gamma = E_\gamma(\mathcal S^i, \theta^i, \phi^i)$,
i.e. the average energy of the primary photon, inducing the shower
with the same arrival direction and $\mathcal S$. 
The look-up table for $E_\gamma(\mathcal S, \theta, \phi)$ is built 
using the photon MC set; the dependence on azimuthal angle, $\phi$, 
is relevant for events with $E_\gamma > 10^{19.5}$~eV where 
geomagnetic preshowering is substantial. 
Photon-induced showers are naturally highly fluctuating.
Consequently, the accuracy of the determination of $E_\gamma$ is 
about 50\% at the one sigma level. 
In the present analysis, $E_\gamma$ is used for
event selection only and therefore its fluctuations are well
accounted for in the exposure calculation. The effect of these 
fluctuations is ``lost'' photons~\cite{ec_method}, i.e. the photons
with reconstructed energy below the energy cut.   
This will be estimated in Section~\ref{sec_Exposure}.

We imposed the following requirements on both the data and MC events:
\begin{enumerate}
\item The shower core is inside the array boundary with the distance to
  the boundary larger than 1200\,m;
\item Zenith angle cut: $45^\circ < \theta < 60^\circ$;
\item The number of scintillator detectors triggered is $\ge$7;
\item The joint fit quality cut, $\chi^2/$d.o.f.$<5$;
\item $\mathcal S$ cut: $E_\gamma(\mathcal S_{obs}^i, \theta^i, \phi^i) > 10^{19}$\,eV or
  $E_\gamma > 10^{19.5}$~eV depending on the energy region discussed
  (the second variant is used for both $E_0 = 10^{19.5}$ and $E_0 = 10^{20}$\,eV).
\end{enumerate}

The cuts determine a photon detection efficiency which is greater than
50\% for showers induced by primary photons with energy above $10^{19}$\,eV. 
The calculation of exposure is given in Section~\ref{sec_Exposure}. 
The resulting data set contains 877 events with $E_\gamma > 10^{19}$\,eV and 
$45^\circ < \theta < 60^\circ$ which we used for our photon search.

\begin{figure*}
\begin{tabular}{cc}
\includegraphics[height=0.45\textwidth,angle=-90]{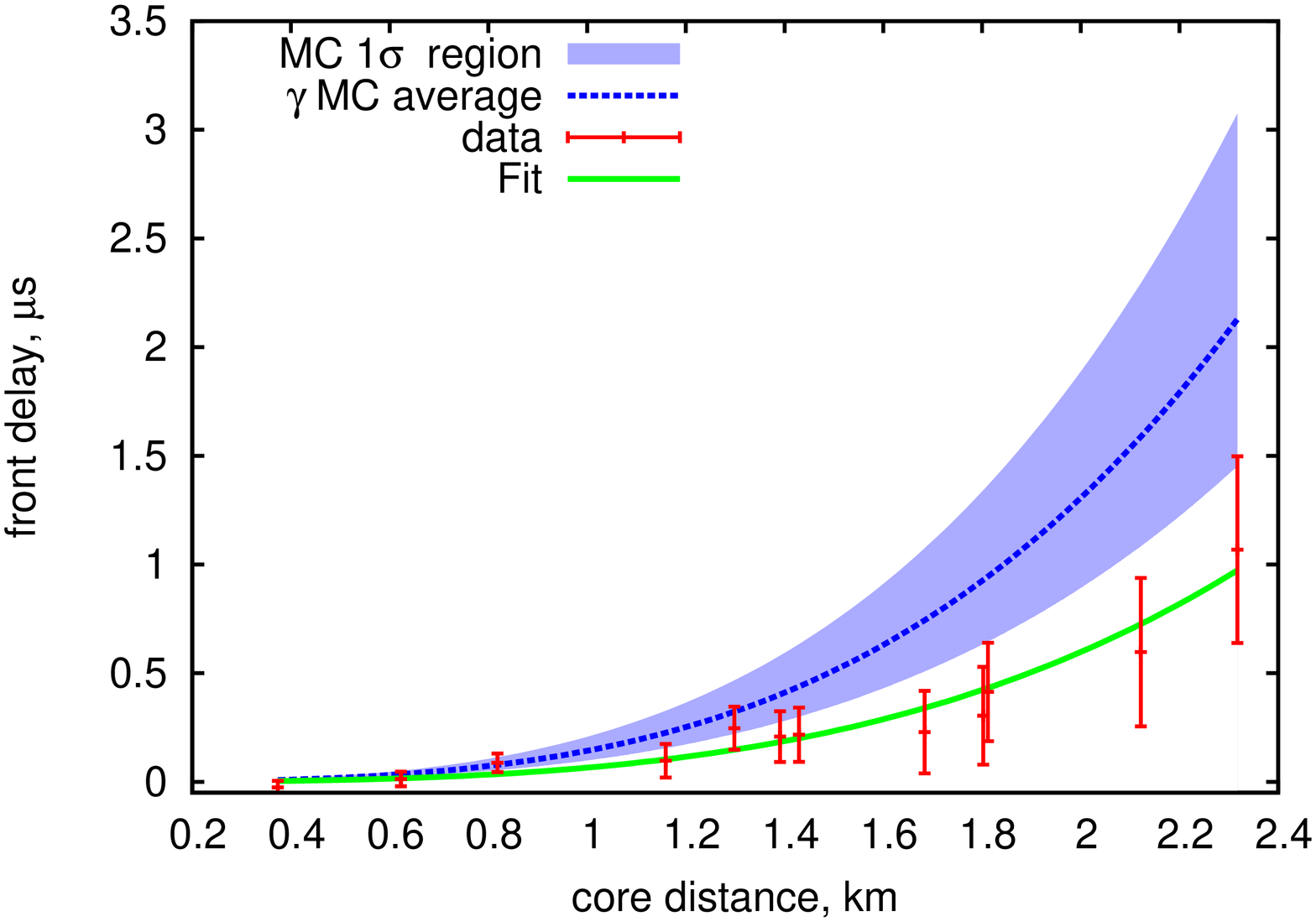} &
\includegraphics[height=0.45\textwidth,angle=-90]{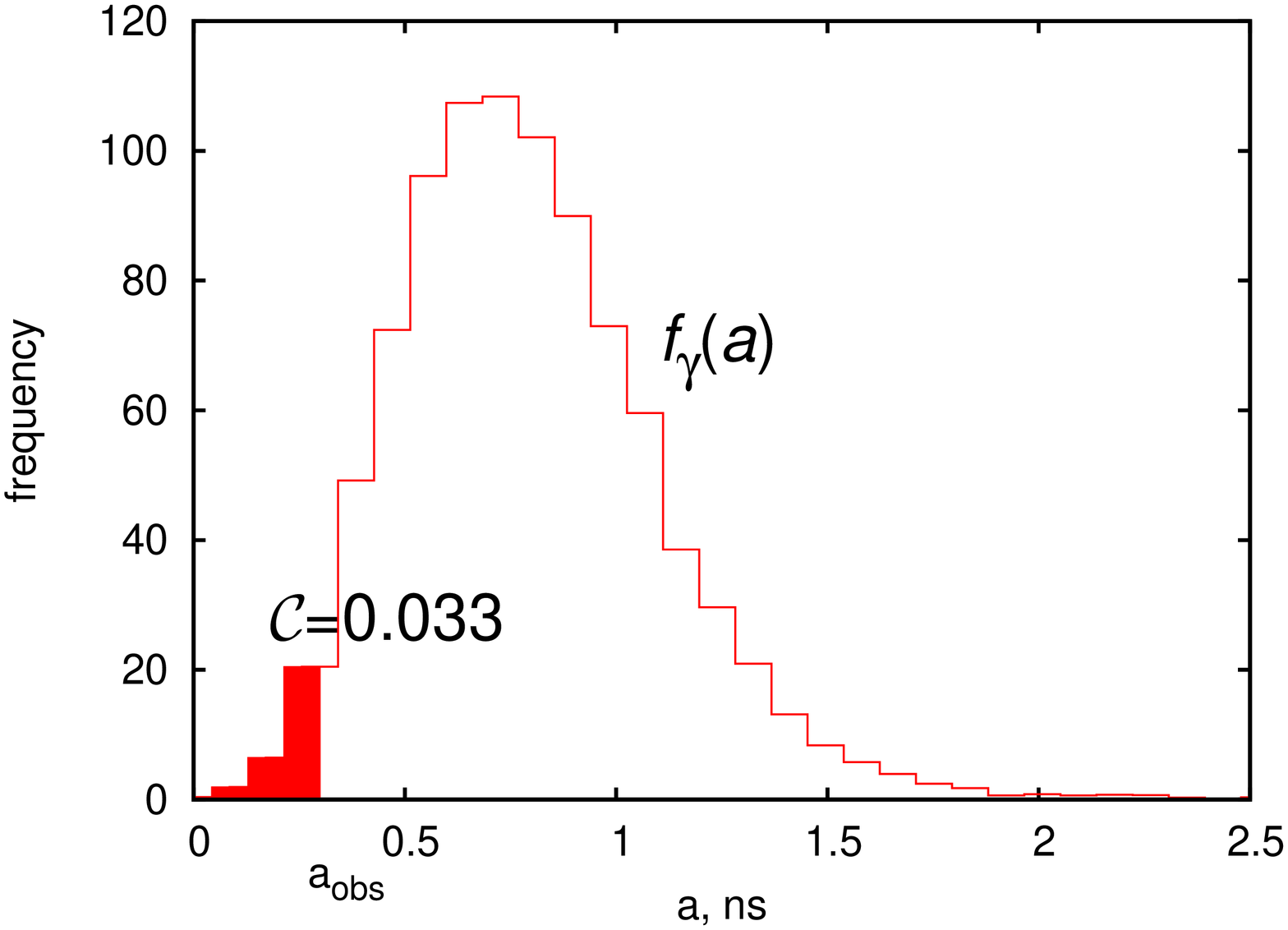} 
\end{tabular}
\caption{\label{fig:ex}
Left: Fit of the shower front for an event 
(2008-08-13 14:02:01, $\theta=$53.6$^\circ$, $ E_\gamma=$ $1.29\times 10^{19}$\,eV, $\mathcal
C$=0.033) compared to average over all photon MC events with the same
zenith angle and $\mathcal S$. 68\% of MC events have a shower front
delay within the 
shaded $1\sigma$ region.
The front delay is counted from the plane front arrival time. 
Right: $f_\gamma(a)$ for the same event; $a_{obs}$ --
observed value of curvature. 
The filled region indicates MC events with
curvature smaller than $a_{obs}$  ($3.3\%$ of MC events).}
\end{figure*}

\section{Method}
\label{sec_Method}

To estimate the flux limit, we used an event-by-event
method~\cite{ec_method}.  The Linsley curvature parameter ``$a$'' is
used as a C-observable and $\mathcal S \equiv S_{800}$ is used as an
E-observable.  For each real event, ``$i$'', we estimate the pair of
parameters ($\mathcal S_{obs}^i$, $a_{obs}^i$) and the arrival
direction ($\theta^i$, $\phi^i$) from the fit of shower front geometry
and LDF.  Histograms of Linsley curvature are shown in
Figure~\ref{fig:a}.  Note that both data and MC distributions show
smallest variance in the region $45^\circ < \theta < 60^\circ$. The latter
motivates the selection of zenith angle range for the further procedure.

We selected simulated gamma-induced showers compatible with the observed
$\theta^i$, $\phi^i$ and $\mathcal S_{obs}^i$ and calculate the
curvature distribution of the simulated photon showers,
$f^i_{\gamma}(a)$, as discussed in Reference~\cite{ec_method}. 
For each event, we determined the percentile rank of 
Linsley parameter, $a$, for photon primaries
$$\mathcal C^i = \int\limits_{-\infty}^{a_{obs}^i} f^i_{\gamma}(a)
da\,,$$ which is the value of the integral probability distribution
function at the observed curvature. 
The shower front fit, $f_\gamma(a)$, and $\mathcal C$ for 
one of the events is shown in Figure~\ref{fig:ex}.

\begin{figure}[h]
{
\includegraphics[width=0.8\columnwidth]{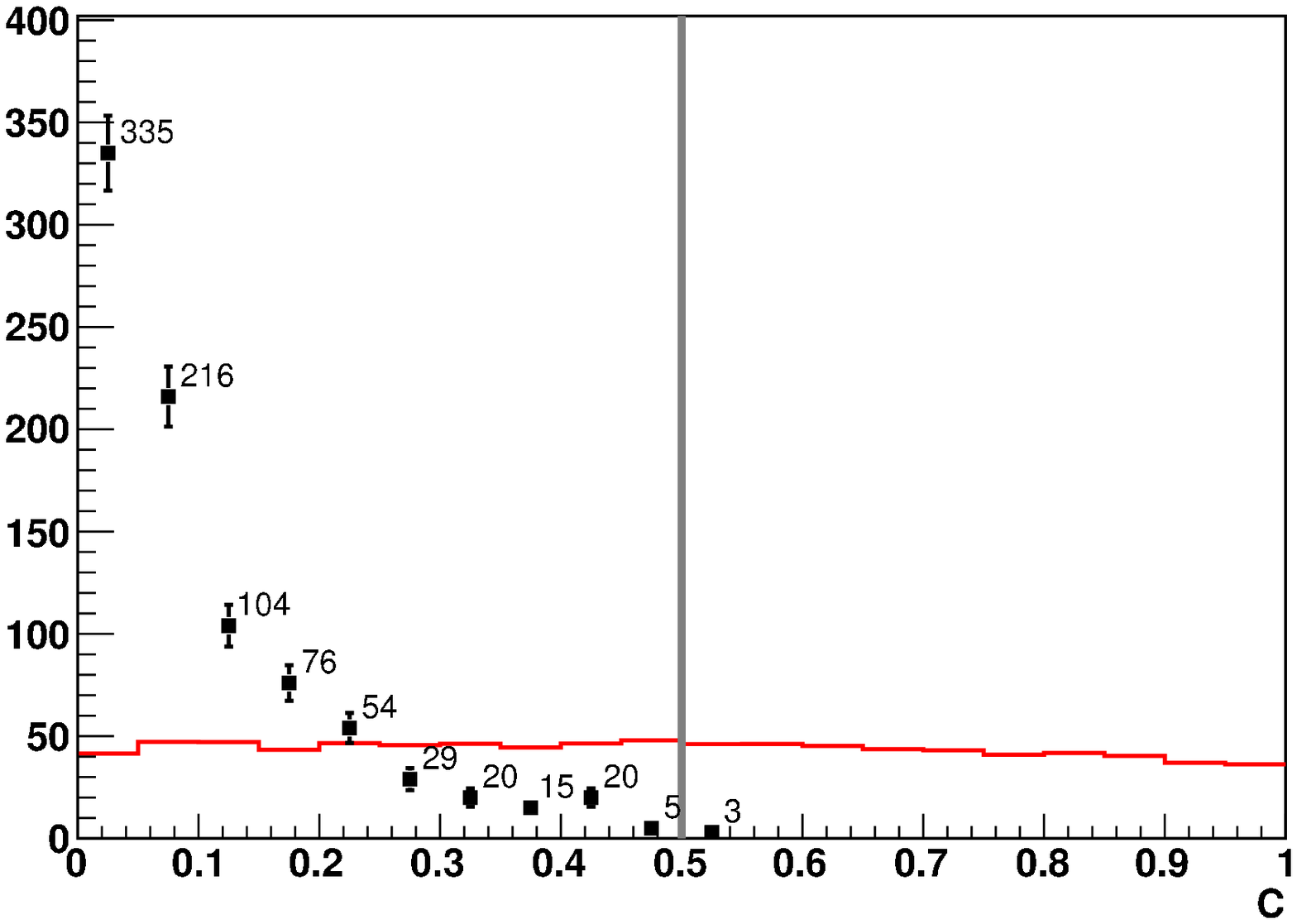} 
}
\caption{\label{fig:C} $\mathcal C$ distribution for the data set
$E_\gamma > 10^{19}$\,eV, $45^\circ < \theta < 60^\circ$. The black points
show the data and the red line indicates the photon MC generated 
with an $E^{-2}$ spectrum. 
The MC photon median is represented by the vertical gray line. }
\end{figure}

The distribution of $\mathcal C$ for the data and MC is shown in Figure~\ref{fig:C}. 
Although the distribution of $f^i_\gamma(a)$ varies with energy and 
arrival direction, $\mathcal C^i$ for gamma-ray primaries would be distributed 
between 0 and 1 uniformly by definition~\footnote{Due to limited MC
statistics $\mathcal C$ distribution for photon MC events differ from uniform. 
The deviations do not exceed 5\%, see
Figure~\ref{fig:C}.}. On the other hand, the actual distribution of
$\mathcal C^i$ in the data is strongly non-uniform (most of the events
have $\mathcal C^i$ below 0.5).

Since the simulations of hadron-induced showers depend strongly on the
hadronic interaction model, we do not use the hadronic showers
simulations in calculation of the photon limit.

Suppose that the integral flux of primary photons over a given
energy range is $F_{\gamma}$. 
Then we expect to detect 
\begin{equation}\label{nbareq}
\bar n (F_{\gamma}) = (1-\lambda) F_{\gamma} A_{geom}
\end{equation}
photon
events on average, where $A_{geom}$ is the geometrical exposure of the
experiment for a given data set and $\lambda$ is the fraction of ``lost''
photons (i.e. photons with primary energies within the interesting
region which failed to enter the data set due to triggering efficiency 
and cuts).

We calculate an upper limit on the primary photon flux
based on the idea that photons satisfy a uniform distribution from 0
to 1 of the variable $\mathcal{C}$. 
To do this, we examine all possible combinations of n events from 
the data set, where n covers the range from 3 to some large value~$M$.  
We compare each combination to a uniform $\mathcal{C}$
distribution using the Smirnov-Cramer-von Mises test~\cite{SCMtest},
and let $\mathcal P(n)$ be the largest probability found in this way. 
By definition of the test $\mathcal P(0)\equiv \mathcal P(1) \equiv \mathcal
P(2) \equiv 1$ and we assume $M=100$ (for which all probabilities
vanish in the considered cases). 
See Appendix~B for a description of the Smirnov-Cramer-von Mises test. 
To constrain the flux $F_{\gamma}$ at the confidence level $\xi$, we require
\begin{equation}\label{CLeq}
\sum_{n=0}^{M} \mathcal P(n) W(n,\bar n(F_{\gamma})) < 1 - \xi\,,
\end{equation}
where $W(n,\bar n)$ is the Poisson  probability of finding $n$ events
when the mean is $\bar n$. 
To constrain the flux at the 95\% confidence level (CL) we set $\xi=0.95$
and find $\bar n$ from equation \eqref{CLeq}. The 
upper limit on the flux follows from equations \eqref{nbareq},
\begin{equation}
F_\gamma < \frac{\bar n}{(1-\lambda) A_{geom}}\,.
\end{equation}

This method does not require any assumptions about hadron-induced
showers and does not require the C-observable to be strongly
discriminating (like the muon density used in~\cite{Ylim,Ylim18,A+Y}).

\section{Exposure}
\label{sec_Exposure}

The geometrical exposure for the SD observation period with $45^\circ <
\theta < 60^\circ$ and boundary cut is
\begin{equation}
A_{geom} = 1286~\mbox{km}^{2}\,\mbox{sr\,yr}\,.
\end{equation}

The fraction of ``lost'' photons is calculated using a photon MC set
generated with an $E^{-2}$ spectrum.  
The values of $(1-\lambda)$ after
consecutive application of cuts are shown in Table~\ref{tab:lost}.

\begin{table}
\begin{center}
\begin{tabular}{|c|c|c|c|}
\hline
~   & \multicolumn{3}{c|}{$E_0$, eV} \\
Cut & $~10^{19}~$ & $~10^{19.5}~$ & $~10^{20}~$ \\
\hline
$n_{det}\ge 7$ & 72\% & 94\% & 97\%\\
$\chi^2$/d.o.f. $<5$ & 68\% & 89\% & 95\%\\
$\mathcal S$ cut & 57\% & 70\% & 95\%\\
\hline
Total: & 57\% & 70\% & 95\%\\
\hline
\end{tabular}
\end{center}
\caption{ \label{tab:lost} Relative exposure of TA SD $(1-\lambda)$ to
  photons after consecutive application of the cuts.  }
\end{table}

\begin{figure}
\begin{center}
  \includegraphics[width=0.90\columnwidth]{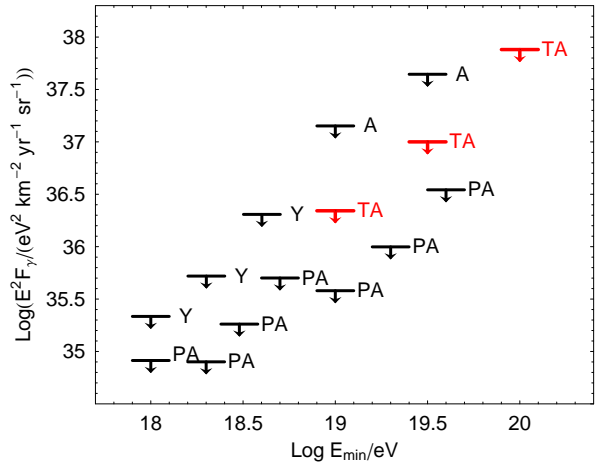} 
\end{center}
\caption{\label{fig:result}
Photon flux limits of the present work~(TA) compared to the previous
limits by AGASA~(A)~\cite{AGASA_1stlim}, Yakutsk~(Y)~\cite{Ylim18} and
Pierre Auger Observatory~(PA)~\cite{Auger_sdlim,Auger_hyblim2}.
}
\end{figure}

\section{Results}
Using the statistical method (Section~\ref{sec_Method}) we arrive at
the following results:
\begin{align*}
\bar n &< 14.1~\mbox{(95\%\,CL)},~E_\gamma > 10^{19}\,\mbox{eV}\,,\\
\bar n &< 8.7~\mbox{(95\%\,CL)},~E_\gamma > 10^{19.5}\,\mbox{eV}\,,\\
\bar n &< 8.7~\mbox{(95\%\,CL)},~E_\gamma > 10^{20}\,\mbox{eV}\,.
\end{align*}
\begin{align*}
F_\gamma &< 1.9\times10^{-2}~\mbox{km}^{-2}\mbox{sr}^{-1}\mbox{yr}^{-1}~\mbox{(95\%\,CL)},\,E_\gamma > 10^{19}\,\mbox{eV}\,,\\
F_\gamma &< 0.97\times10^{-2}~\mbox{km}^{-2}\mbox{sr}^{-1}\mbox{yr}^{-1}~\mbox{(95\%\,CL)},\,E_\gamma > 10^{19.5}\,\mbox{eV}\,,\\
F_\gamma &< 0.71\times10^{-2}~\mbox{km}^{-2}\mbox{sr}^{-1}\mbox{yr}^{-1}~\mbox{(95\%\,CL)},\,E_\gamma > 10^{20}\,\mbox{eV}\,.
\end{align*}

These photon limits are shown along with the results of the other 
experiments in Figure~\ref{fig:result}.

We obtain photon fraction limits by dividing the corresponding flux limits 
by the integral flux of the Telescope Array SD spectrum~\cite{TAspec}:
\begin{align*}
\varepsilon_\gamma &< 6.2\%~\mbox{(95\% CL)},~E_\gamma > 10^{19}~\mbox{eV}\,,\\
\varepsilon_\gamma &< 28.5\%~\mbox{(95\% CL)},~E_\gamma > 10^{19.5}~\mbox{eV}\,.
\end{align*}

The limits strongly constrain the top-down models of the origin of cosmic
rays, see~\cite{WG} for discussion.


Next, we compare the results of the event-by-event method with the results of
the simpler ``photon median'' method~\cite{Auger_sdlim}. 
In the latter, the events having curvature greater than the median photon 
curvature are identified as photon candidates. 
This criteria corresponds to $\mathcal C > 0.5$. We
we observe three candidate events with energy greater than $10^{19}$~eV 
(see Figure~\ref{fig:C})  and no candidate events above $10^{19.5}$~eV. 
This corresponds to a 95\% Poisson confidence limit of 
$\bar n/2 < 8.25$ and $\bar n/2 < 3.09$. The  
flux limits are $F_\gamma < 2.3\times10^{-2}$, $F_\gamma < 0.69\times10^{-2}$ 
and $F_\gamma < 0.51\times10^{-2}~\mbox{km}^{-2}\mbox{sr}^{-1}\mbox{yr}^{-1}$ 
for $E_0 = 10^{19}$, $10^{19.5}$ and $10^{20}$~eV correspondingly. 
The limits using the two methods are in mutual agreement.

Finally we discuss how the result depends on the assumption of the
$E^{-2}$ primary photon spectrum. We repeated the analysis with varied
spectral index and for $E>10^{19}$~eV arrived at $F_\gamma <
2.2\times10^{-2}$ and $F_\gamma <
1.8\times10^{-2}\,\mbox{km}^{-2}\mbox{sr}^{-1}\mbox{yr}^{-1}$ for
$E^{-1.5}$ and $E^{-2.5}$ primary spectra correspondingly. The limits
for energy greater than $10^{19.5}$ and $10^{20}\,$eV are less
sensitive to the spectral assumption.

Both the use of plastic scintillators sensitive to photon-induced
showers and the application of event-by-event statistical method
allowed us to put stringent limits on the flux of primary photons with
energies in excess of $10^{19}$~eV with the data obtained during three
years of the TA surface detector operation. The photons propagate
without deflection by magnetic fields and therefore in the case of the
few nearby sources we may not expect an isotropic flux. It worth
mentioning that the limits of this paper are strongest among those
obtained in the northern hemisphere. The result depends neither on the
choice of hadronic interaction model, nor on possible systematics in
the energy determination of hadronic primaries.

\section*{Acknowledgments}
The Telescope Array experiment is supported by the Japan Society for
the Promotion of Science through Grants-in-Aids for Scientific
Research on Specially Promoted Research (21000002) ``Extreme Phenomena
in the Universe Explored by Highest Energy Cosmic Rays'' and for
Scientific Research (19104006), and the Inter-University Research
Program of the Institute for Cosmic Ray Research; by the U.S. National
Science Foundation awards PHY-0307098, PHY-0601915, PHY-0649681,
PHY-0703893, PHY-0758342, PHY-0848320, PHY-1069280, and PHY-1069286;
by the National Research Foundation of Korea (2007-0093860, R32-10130,
2012R1A1A2008381, 2013004883); by the Russian Academy of Sciences, by
the grant of the President of the Russian Federation MK-1170.2013.2,
Dynasty Foundation, RFBR grants 11-02-01528, 13-02-01311 and
13-02-01293 (INR), IISN project No. 4.4509.10 and Belgian Science
Policy under IUAP VII/37 (ULB). The foundations of Dr. Ezekiel R. and
Edna Wattis Dumke, Willard L. Eccles and the George S. and Dolores
Dore Eccles all helped with generous donations. The State of Utah
supported the project through its Economic Development Board, and the
University of Utah through the Office of the Vice President for
Research. The experimental site became available through the
cooperation of the Utah School and Institutional Trust Lands
Administration (SITLA), U.S. Bureau of Land Management, and the
U.S. Air Force. We also wish to thank the people and the officials of
Millard County, Utah for their steadfast and warm support. We
gratefully acknowledge the contributions from the technical staffs of
our home institutions. An allocation of computer time from the Center
for High Performance Computing at the University of Utah is gratefully
acknowledged. The cluster of the Theoretical Division of INR RAS was
used for the numerical part of the work.

\section*{Appendix A. LDF and shower front fit functions}

We perform joint fit of LDF and shower front with 7 free parameters:
$x_{core}$, $y_{core}$, $\theta$, $\phi$, $S_{800}$, $t_0$, $a$.
\begin{align*}
S(r) &= { S_{800}} \times LDF(r) \,,\\
t_0(r) &={ t_0} + t_{plane} + {a}\times 0.67~(1+r/R_L)^{1.5} LDF^{-0.5}(r)\,,
\end{align*}
where $t_{plane}$ is a shower plane delay, $a$ is a Linsley curvature
parameter and the $LDF(r)$ is defined as follows: 
\begin{align*}
LDF(r) &= f(r)/f\left(800\,\mbox{m}\right)\,,\\
f(r) &=  \left(\frac{r}{R_m}\right)^{-1.2}
\left(1+\frac{r}{R_m}\right)^{-(\eta-1.2)}
\left(1+\frac{r^2}{R_1^2}\right)^{-0.6}\,,
\end{align*}
$$R_m = 90\,\mbox{m},~R_1 = 1000\,\mbox{m},~R_L = 30\,\mbox{m},$$
$$\eta = 3.97 - 1.79 \times (\sec(\theta) -1)\,.$$

\section*{Appendix B. Smirnov-Cramer-von Mises ``omega-square'' test implementation}

Let $F(x)$ be theoretical distribution and $F_n(x)$ -- observed
distribution of $n$ events. We define the distance between
distributions by~\cite{SCMtest}:
$$ \omega^2 = \int\limits_{-\infty}^{\infty} \left( F_n(\mathcal C) -
F(\mathcal C)
\right)^2 d F(\mathcal C) \;.$$

If $\mathcal C_1,\mathcal C_2,\ldots,\mathcal C_n$ is a set of
observed values in increasing order, $\omega^2$ may be rewritten in
the following form:

$$ n\, \omega^2 =\frac{1}{12n} +\sum\limits_{i=1}^n \left(\frac{2i-1}{2n} -
F(\mathcal C_i) \right)^2\;.$$

In this paper, we compare the distribution of an event subset with uniform
distribution $U(0,1)$. Therefore $F(\mathcal C_i) = \mathcal
C_i$ and we have:
$$ n\, \omega^2 =\frac{1}{12n} +\sum\limits_{i=1}^n \left(\frac{2i-1}{2n} -
\mathcal C_i \right)^2\;.$$
 The required maximization of the probability over subsets is therefore
reduced to the selection of $n$ different events minimizing the above
sum. The latter may be done with a fast iterative procedure.

\end{document}